# Neural Networks Compensation of Systems with Multi-segment Piecewise Linear Nonlinearities

Jun Oh JANG

Department of Software Engineering

Uiduk University, Gyeongju, 380004, South Korea

Tel. : +82-054-760-1624, Email : jojang@uu.ac.kr

Abstract: A neural networks (NN) compensator is designed for systems with multi-segment piecewise-linear nonlinearities. The compensator uses the back stepping technique with NN for inverting the multi-segment piecewise-linear nonlinearities in the feedforward path. This scheme provides a general procedure for determining the dynamic pre-inversion of an invertible dynamic system using NN. A tuning algorithm is presented for the NN compensator which yields a stable closed-loop system. In the case of nonlinear stability proofs, the tracking error is small. It is noted that PI controller without NN compensation requires much higher gain to achieve same performance. It is also difficult to ensure the stability of such highly nonlinear systems using only PI controllers. Using NN compensation, stability of the system is proven, and tracking errors can be arbitrarily kept small by increasing the gain. The NN weight errors are basically bounded in terms of input weight and hidden weight. Simulation results show the effectiveness of the piecewise linear NN compensator in the system. This scheme is applicable to xy table-like servo system and shows neural network stability proofs. In addition, the NN piecewise linear nonlinearity compensation can be further and applied to backlash, hysteresis, and another actuator nonlinear compensation.

## 1. Introduction

Piecewise linear nonlinearity often occurs in real control systems. In these systems, the actuator operates with different gains in different areas of the control signal. A simple example of a piecewise linear system is the heating and cooling process [1]. Here, "thermal gain" is different from "cooling gain". Another example is nonlinear servosystem [2], where the system has a high gain for a large input and a low gain for a small input. A more complex example of a piecewise linear system is a system having different linear gains for different computational regions. These system models are widely used to approximate systems with general actuator nonlinear characteristics.

The adaptive control of the piecewise linear system has been studied for two segment settings in [3] and [4] that switch the actuator gain switch from one value to another when the control input passes zero. Output regulation of discrete time piecewise linear nonlinearities was proposed in [5]. Recently, several strictly induced adaptive schemes have been provided for actuator nonlinearity compensation in major tasks [6]. Backlash compensation using dynamic inversion is presented in [7] and [8] for discrete time, where fuzzy logic is used to cancel the inversion error.  The hysteresis compensation of systems using neural networks was proposed in [9].

In this paper, we design an NN compensation scheme for systems with multi-segment piecewise linear nonlinearity. The actuator gain may switch several values in different operating regions. Strict design procedures are provided with proof that a PI tracking loop using adaptive NN systems occurs in a feedforward loop for multi-segment piecewise linear nonlinearities compensation. The proposed method can be applied to compensation for large multi-segment piecewise linear nonlinearity classes. The author derives actual limitations on tracking error from the analysis of the tracking error dynamics and investigates the performance of an NN compensator in the system through computer simulations.

## 2. Neural Networks

NN have been widely used in feedback control systems [10-15]. Most applications are temporary without proving their stability. The stability proofs present almost always depends on the universal approximation property for NN [16, 17]. The three layers NN in Figure 1 consist of an input layer, a hidden layer, and an output layer. There are $L$ neurons in the hidden layer, and $m$ neurons in the output layer. The multilayer NN is a nonlinear mapping from the input space $R^n$ to the output space $R^m$.

The NN output $y$ is a vector with $m$ components determined by $n$ components of the input vector $x$ by the equation

$$y_i = \sum_{k=1}^{L}[w_{ik}\sigma(\sum_{j=1}^{n}v_{kj}x_j + v_{k0}) + w_{i0}] \; ; \quad i = 1,2,\cdots,m \tag{1}$$

where $\sigma(\cdot)$ are the hyperbolic tangent function, $v_{kj}$, the connection weights from input to hidden layer, $w_{ik}$, connection weights from hidden to output layer. The threshold offsets are displayed as $v_{k0}$, $w_{i0}$. By collecting all the NN weights, $v_{kj}, w_{ik}$ into matrices $V^T, W^T$, the NN equation can be written in vector as follows

$$y = W^T \sigma(V^T x) \tag{2}$$

The threshold is included as the first column of the weight matrices $W^T$, $V^T$; to accommodate this, the vector $x$ and $\sigma(\cdot)$ must be increased by placing '1' as the first element(e.g. $x = [1 \; x_1 \; x_2 \; \cdots \; x_n]^T$). In this equation, to represent (1) it has sufficient generality if $\sigma(\cdot)$ is taken as a diagonal function from $R^L$ to $R^L$, that is, $\sigma(z) = diag\{\sigma(z_k)\}$ for a vector $z = [z_1 \; z_2 \; \cdots \; z_L]^T \in R^L$. For notational convenience, all matrices of weights are defined by as follows

$$Z = \begin{bmatrix} W \\ V \end{bmatrix}. \tag{3}$$

Many well-known results show that a sufficiently smooth function $\bar{y}$ can be approximated arbitrarily closely in a compact set using a three-layer NN with an appropriate weights, i.e.

$$\bar{y} = W^T \sigma(V^T x) + \varepsilon(x) \tag{4}$$

where $\varepsilon(x)$ is the NN approximation error, and $\| \varepsilon(x) \| \leq \varepsilon_N$ on a compact set $S$ [18, 19]. The approximating weights $V$ and $W$ are ideal target weights, and it is supposed that they are limited such as $\|V\|_F \leq V_M, \|W\|_F \leq W_M$, or $\|Z\|_F \leq Z_M$.

## 3. Piecewise Linear Nonlinearities

In this section author presents the piecewise-linear nonlinearities model and its inverse model. An implementation of the piecewise-linear nonlinearities inverse is given by for developing our NN compensation scheme for systems with piecewise-linear nonlinearities in the next section. Nonlinearity compensation is done using dynamic inversions, where the NN is used for the dynamic inversion compensation [20].

The model of the piecewise-linear nonlinearities is shown in Fig. 2. The piecewise-linear nonlinearities characteristic $P(\cdot)$ with input $u(t)$ and output $T(t)$ is as follows :

$$T = P(u(t)) = \begin{cases} m_{r1}u(t), & \text{if } 0 < u(t) \leq u(t) \\ m_{r2}(u(t) - u_r) + m_{r1}u_r, & \text{if } u(t) > u_r \\ m_{l1}u(t), & \text{if } u_l \leq u(t) < 0 \\ m_{l2}(u(t) - u_l) + m_{l1}u_l, & \text{if } u(t) < u_l \end{cases} \quad (5)$$

where $m_{r1}, m_{r2}, m_{l1}, m_{l2}, u_r, u_l$ are constant parameters as are $T_r = m_{r1} \cdot u_r, T_l = m_{l1} \cdot u_l$. In order to cancel the effect of piecewise linear nonlinearities in the system, the NN precompensator needs to generate inverse of the piecewise linear nonlinearities. Fig. 3 shows the piecewise linear nonlinearities inverse function. The dynamics of the NN compensator is given by

$$\dot{u}(t) = H_{inv}(u, u_d, \dot{u}_d). \quad (6)$$

The piecewise linear nonlinearities inverse characteristic in Figure 3 can be decomposed into two functions. In other words, it is a direct feedforward term and the additional modified inverse term as shown in Figure 4. This decomposition can design a compensator with a better structure when using NN in the feedforward path.

## 4. NN piecewise Linear Nonlinearities Compensation of Systems

The NN compensator is designed using the back stepping technique [21]. In this section the author will show how to tune or learn weights of NN on-line so that the tracking errors are small and the NN weights are bounded. Assumed that the actuator output $T(t)$ is measured.

The dynamics of the system with no vibratory modes can be described as:

$$M\ddot{\theta} + B\dot{\theta} + T_f + T_d = T \quad (7)$$

where $\dot{\theta}$ is the velocity of the system, $M$ is the mass, $B$ is the damping, $T_f$ is the nonlinear friction torque, $T_d$ is the bounded unknown disturbance, and $T$ is the control input. It is assumed that $|T_d| < \tau_M$, with $\tau_M$, a known positive constant. Given the reference signal $\dot{\theta}_d$, the tracking error is expressed by $e = \dot{\theta}_d - \dot{\theta}$. The reference signal is bounded so that $|\dot{\theta}_d| < \Theta_d, \Theta_d$ is a known constant. Differentiating tracking error and

using (7), the system dynamics may be described in terms of the tracking error as:

$$M\dot{e} = -Be - T + f(x) + T_d \tag{8}$$

where the nonlinear plant function is defined as

$$f(x) = M\ddot{\theta}_d + B\dot{\theta}_d + T_f. \tag{9}$$

The term $x$ includes all the time signals required for calculation $f(\cdot)$, and can be defined as follows $x \equiv [\dot{\theta}_d \quad \ddot{\theta}_d]^T$. The function $f(x)$ contains all potentially unknown functions, except for $M$, $B$ shown in (8) – these latter is offset by the proof of stability.

A robust compensation scheme for unknown terms in $f(x)$ is provided by selecting the tracking controller

$$T_{des} = K_f e + \hat{f} - v_1 \tag{10}$$

with $\hat{f}(x)$, an estimate for the nonlinear terms $f(x)$, $v_1(t)$ a robustifying term, and $K_f > 0$. Fig. 5 shows the control structure implied by this scheme. The proposed controller has a Proportional-Integral (PI) tracking loop with gains $K_f e = K_I \int e + K_p e$ where the piecewise linear effect is ameliorated by a piecewise linear compensator. It is assumed that the nonlinear function $f(x)$ is unknown. But a fixed estimate $\hat{f}(x)$ is assumed to know the function estimation error, $\tilde{f}(x) = f(x) - \hat{f}(x)$, satisfies $|\tilde{f}(x)| \leq f_M(x)$, for some known bounding function $f_M(x)$.

The following theorem is the first step in the backstepping design, showing that the desired control law (10) will keeps the tracking error small.

*Theorem 1*: Given the system (7), select the tracking control law (10). Choose the robustifying signal $v_1$ as

$$v_1(t) = -(f_M(x) + \tau_M)\frac{e}{|e|}. \tag{11}$$

Then the tracking error is bounded and it can be kept as desired by increasing the gains $K_f$.

Proof: Select the Layapnov function candidate

$$L_1 = \frac{1}{2}Me^2. \tag{12}$$

Differentiating $L_1$ and using (8) and the assumption $|\dot{M}| = 0$ yields

$$\dot{L}_1 = e(-Be + f + T_d - T_{des}). \tag{13}$$

Applying the tracking control law (10) one has

$$\begin{aligned}\dot{L}_1 &= e(-Be + f + T_d - K_f e - \hat{f} + v_1) \\ &= -(K_f + B)e^2 + e(\tilde{f} + T_d + v_1)\end{aligned} \tag{14}$$

$$\dot{L}_1 = -(K_f + B)e^2 + e\{\tilde{f} + T_d - (f_M + \tau_M)\frac{e}{|e|}\}. \tag{15}$$

Equation (15) can be bounded as

$$\dot{L}_1 \leq -(K_f + B)|e|^2 - |e|(f_M + \tau_M) + |e|\,\|\tilde{f} + T_d\|. \tag{16}$$

One can conclude that $\dot{L}_1$ is guaranteed negative for as long as $|e| \neq 0$.

Theorem 1 gives a control law that guarantees stability in terms of the tracking error. In the presence of the unknown piecewise linear nonlinearity, the desired value and actual value of the control signal $T$ will be different. Following the idea of dynamic inversion where NN is used for compensation of the inversion error, originally given by Calise et al. [21], the author gives a rigorous analysis of the closed loop system stability.

The actuator output provided by (10) is a desirable ideal signal. To find the overall system error dynamics, define the error between the desired actuator output and the actual actuator output as follows.

$$\tilde{T} = T_{des} - T. \tag{17}$$

Differentiating one has

$$\begin{aligned}\dot{\tilde{T}} &= \dot{T}_{des} - \dot{T} \\ &= \dot{T}_{des} - H(T,u,\dot{u})\end{aligned} \tag{18}$$

which together with (8) and involving (10) represent the complete system error dynamics.

The dynamics of the piecewise linear nonlinearity can be written as

$$\dot{T} = \varphi \tag{19}$$

$$\varphi = P(T,u,\dot{u}) \tag{20}$$

where $\varphi(t)$ is pseudo-control input. In the case of known piecewise linear, the ideal piecewise linear inverse is given by

$$\dot{u} = P^{-1}(u,T,\varphi). \tag{21}$$

Since the piecewise linear and therefore its inverse are not known, so only the piecewise linear inverse can be approximated as

$$\dot{\hat{u}} = \hat{P}^{-1}(\hat{u},T,\hat{\varphi}). \tag{22}$$

The dynamics of the piecewise linear nonlinearities can now be written as

$$\begin{aligned}\dot{T} &= P(T,\hat{u},\dot{\hat{u}}) \\ &= \hat{P}(T,\hat{u},\dot{\hat{u}}) + \tilde{P}(T,\hat{u},\dot{\hat{u}}), \\ &= \hat{\varphi} + \tilde{P}(T,\hat{u},\dot{\hat{u}})\end{aligned} \tag{23}$$

where $\hat{\varphi} = \hat{P}(T,\hat{u},\dot{\hat{u}})$ and therefore its inverse $\dot{\hat{u}} = \hat{P}^{-1}(T,\hat{u},\dot{\hat{u}})$. The unknown function $\tilde{P}(T,\hat{u},\dot{\hat{u}})$, which represents the piecewise linear inversion error, will be approximated using NN.

Based on the NN approximation property, the piecewise linear inversion error can be represented as

$$\tilde{P}(T,\hat{u},\dot{\hat{u}}) = W^T \sigma(V^T x_{nn}) + \varepsilon(x_{nn}), \tag{24}$$

where the NN input vector is chosen as $x_{nn} = [1 \quad e \quad \dot{\theta}_d \quad \tilde{T} \quad T \quad \|\hat{Z}\|_F]^T$, and $\varepsilon$ represents the NN approximation error.

$\hat{V}, \hat{W}$ is defined as an estimates of the ideal NN weights provided by the NN tuning algorithms. Define the weight estimation error as

$$\tilde{V} = V - \hat{V}, \quad \tilde{W} = W - \hat{W}, \quad \tilde{Z} = Z - \hat{Z}, \tag{25}$$

and the hidden layer output error for a given $x_{nn}$ as

$$\tilde{\sigma} = \sigma - \hat{\sigma} = \sigma(V^T x_{nn}) - \sigma(\hat{V}^T x_{nn}). \tag{26}$$

For a stable closed-loop system design through piecewise linear nonlinearities compensation, one selects a nominal piecewise linear inverse $\dot{\hat{u}} = \hat{\varphi}$ and pseudo-control input $\hat{\varphi}$ as

$$\hat{\varphi} = K_b \tilde{T} + \dot{T}_{des} - \hat{W}^T \sigma(\hat{V}^T x_{nn}) - v_2 \tag{27}$$

where $v_2(t)$ is a robustifying term detailed later.

The closed loop system with the NN compensator is shown in Fig. 5. The proposed NN compensation scheme is in accordance with the piecewise linear nonlinearities inverse decomposition in Fig. 4. The exact piecewise linear nonlinearities inverse consists of a direct feed term and the error term in Fig. 4(b) which is estimated by NN.

Using the proposed controller (27), the error dynamics (18) can be written as

$$\begin{aligned}\dot{\tilde{T}} &= \dot{T}_{des} - \hat{\varphi} - \tilde{P}(T, \hat{u}, \dot{\hat{u}}) \\ &= -K_b \tilde{T} + \hat{W}^T \sigma(\hat{V}^T x_{nn}) + v_2 - W^T \sigma(V^T x_{nn}) - \varepsilon(x_{nn})\end{aligned}. \tag{28}$$

More precisely, the error dynamics can be described as :

$$\begin{aligned}\dot{\tilde{T}} &= -K_b \tilde{T} + \hat{W}^T \sigma(\hat{V}^T x_{nn}) + W^T \sigma(\hat{V}^T x_{nn}) - W^T \sigma(\hat{V}^T x_{nn}) \\ &\quad + v_2 - W^T \sigma(V^T x_{nn}) - \varepsilon(x_{nn})\end{aligned}, \tag{29}$$

$$\dot{\tilde{T}} = -K_b \tilde{T} - \tilde{W}^T \hat{\sigma} - W^T \tilde{\sigma} + v_2 - \varepsilon(x_{nn}), \tag{30}$$

with the NN input is bounded by

$$\|x_{nn}\| \leq c_0 + |e| + \Theta_d + |\tilde{T}| + c_1 \|\tilde{Z}\|_F. \tag{31}$$

The following theorem shows how to adjust the neural network weights so the tracking error $e(t)$ and $\tilde{T}(t)$ achieve small values while the NN weights $\hat{V}$, $\hat{W}$ are close to $V, W$, i.e., the weight estimation errors defined by (25) are bounded.

*Theorem* 2. Bound the desired trajectories. The control input is selected as (27). Choose the robustifying signal $v_2$ as

$$v_2 = -K_{Z_1}(\|\hat{Z}\|_F + Z_M)(\tilde{T} + |e|\frac{\tilde{T}}{|\tilde{T}|}) - K_{Z_2}|e|\frac{\tilde{T}}{|\tilde{T}|} - K_{Z_3}(\|\hat{Z}\|_F + Z_M)^2 \frac{\tilde{T}}{|\tilde{T}|}, \tag{32}$$

where $K_{Z_1} > \sqrt{L}$, $K_{Z_2} > 1$, and $K_{Z_3} > c_1 \sqrt{L}$, where $L$ is the number of hidden layer nodes. Let the estimated NN weights be given by the NN tuning algorithm

$$\dot{\hat{V}} = -Q |\tilde{T}| x_{nn} \hat{\sigma}^T - kQ |\tilde{T}| \hat{V},$$

$$\dot{\hat{W}} = -S\hat{\sigma}\tilde{T} - kS |\tilde{T}| \hat{W}, \qquad (33)$$

with any constant matrices $S = S^T > 0$, $Q = Q^T > 0$, and $k > 0$ small scalar design parameter. Then the tracking error $e(t)$, error $\tilde{T}(t)$ and NN weight estimates $\hat{V}, \hat{W}$ are bounded, with bounds provided by Eqs. (49) and (50). In addition, the error $\tilde{T}(t)$ can be arbitrarily made small by increasing the gain $K_b$.

*Proof*: Select the Lyapnov function candidate

$$L = L_1 + \frac{1}{2}\tilde{T}^2 + \frac{1}{2}tr(\tilde{W}^T S^{-1} \tilde{W}) + \frac{1}{2}tr(\tilde{V}^T Q^{-1} \tilde{V}) \qquad (34)$$

which weights both errors $e(t)$ and $\tilde{T}(t)$, and NN weights estimation errors. Taking derivative

$$\dot{L} = \dot{L}_1 + \tilde{T}\dot{\tilde{T}} + tr(\tilde{W}^T S^{-1} \dot{\tilde{W}}) + tr(\tilde{V}^T Q^{-1} \dot{\tilde{V}}), \qquad (35)$$

and using (8), (30) one has

$$\dot{L} = e(-Be - T + f + T_d) + \tilde{T}(-K_b\tilde{T} - \tilde{W}^T \hat{\sigma} - W^T \tilde{\sigma} + v_2 - \varepsilon(x))$$
$$+ tr(\tilde{W}^T S^{-1} \dot{\tilde{W}}) + tr(\tilde{V}^T Q^{-1} \dot{\tilde{V}}) \qquad (36)$$

$$\dot{L} = e(-Be + f + T_d - T_{des}) + e\tilde{T} + \tilde{T}(-K_b\tilde{T} - \tilde{W}^T \hat{\sigma}$$
$$- W^T \tilde{\sigma} + v_2 - \varepsilon(x)) + tr(\tilde{W}^T S^{-1} \dot{\tilde{W}}) + tr(\tilde{V}^T Q^{-1} \dot{\tilde{V}}) \qquad (37)$$

$$\dot{L} = e(-Be + f + T_d - T_{des}) + e\tilde{T} + \tilde{T}(-K_b\tilde{T} + v_2 - \varepsilon(x))$$
$$- \tilde{T}W^T \tilde{\sigma} + tr[\tilde{W}^T (S^{-1}\dot{\tilde{W}} - \hat{\sigma}\tilde{T})] + tr[\tilde{V}^T Q^{-1}\dot{\tilde{V}}] \qquad (38)$$

Applying (10) and tuning rules yields

$$\dot{L} = e(-Be + \tilde{f}(x) + T_d - K_f e + v_1) + e\tilde{T} + \tilde{T}(-K_b\tilde{T} + v_2 - \varepsilon(x))$$
$$- \tilde{T}W^T \tilde{\sigma} + k|\tilde{T}| tr[\tilde{W}^T (W - \tilde{W})] + k|\tilde{T}| tr[\tilde{V}^T (V - \tilde{V})] - |\tilde{T}| tr[\tilde{V}^T x_{nn} \hat{\sigma}^T] \qquad (39)$$

Using (11) and (32), expression (39) can be bounded as

$$\dot{L} \leq -(K_f + B)|e|^2 - |e|(f_M + \tau_M) + |e| \|\tilde{f} + T_d\| + k|\tilde{T}| \|Z\|_F (Z_M - \|\tilde{Z}\|_F)$$
$$- K_b |\tilde{T}|^2 - \tilde{T}\varepsilon(x) + e\tilde{T} - \tilde{T}K_{Z_1}(\|\hat{Z}\|_F + Z_M)(\tilde{T} + |e|\frac{\tilde{T}}{|\tilde{T}|}) - \tilde{T}K_{Z_2}|e|\frac{\tilde{T}}{|\tilde{T}|} \qquad (40)$$
$$- \tilde{T}K_{Z_3}(\|\hat{Z}\|_F + Z_M)^2 \frac{\tilde{T}}{|\tilde{T}|} - \tilde{T}W^T \tilde{\sigma} - |\tilde{T}| tr(\tilde{V}^T x_{nn} \hat{\sigma}^T)$$

Including (31) and applying some norm properties, one can has

$$\dot{L} \leq -(K_f + B)|e|^2 - |e|(f_M + \tau_M) + |e| \|\tilde{f} + T_d\| + k|\tilde{T}| \|Z\|_F (Z_M - \|\tilde{Z}\|_F)$$
$$- K_b |\tilde{T}|^2 + |\tilde{T}\| e| + |\tilde{T}\| \varepsilon(x)| + 2|\tilde{T}|Z_M \sqrt{L} + |\tilde{T}\| \tilde{Z}\|_F |x_{nn}\hat{\sigma}^T| \qquad (41)$$
$$- K_{Z_1} |\tilde{T}|^2 \|\tilde{Z}\|_F - K_{Z_1}|e\|\tilde{T}\| \tilde{Z}\|_F - K_{Z_2}|\tilde{T}\| e| - K_{Z_3}|\tilde{T}\| \tilde{Z}\|_F^2$$

$$\dot{L} \leq -(K_f + B)|e|^2 + kZ_M|\tilde{T}| \|\tilde{Z}\|_F - k|\tilde{T}| \|\tilde{Z}\|_F^2 - K_b|\tilde{T}|^2 + |\tilde{T}\|e| + |\tilde{T}|\varepsilon_N$$
$$+ 2|\tilde{T}|Z_M\sqrt{L} + |\tilde{T}| \|\tilde{Z}\|_F \sqrt{L}(c_0 + |e| + \Theta_d + |\tilde{T}| + c_1\|\tilde{Z}\|_F) \quad (42)$$
$$- K_{Z_1}|\tilde{T}|^2\|\tilde{Z}\|_F - K_{Z_1}|e\|\tilde{T}\| \|\tilde{Z}\|_F - K_{Z_2}|\tilde{T}\|e| - K_{Z_3}|\tilde{T}| \|\tilde{Z}\|_F^2$$

$$\dot{L} \leq -(K_f + B)|e|^2 + kZ_M|\tilde{T}| \|\tilde{Z}\|_F - k|\tilde{T}| \|\tilde{Z}\|_F^2 - K_b|\tilde{T}|^2 + |\tilde{T}\|e|$$
$$+ C_0|\tilde{T}| + C_1|\tilde{T}| \|\tilde{Z}\|_F + \sqrt{L}|\tilde{T}\|e\| \|\tilde{Z}\|_F + \sqrt{L}\|\tilde{Z}\|_F|\tilde{T}|^2 + c_1\sqrt{L}|\tilde{T}| \|\tilde{Z}\|_F^2 \quad (43)$$
$$- K_{Z_1}|\tilde{T}|^2\|\tilde{Z}\|_F - K_{Z_1}|e\|\tilde{T}\| \|\tilde{Z}\|_F - K_{z_2}|\tilde{T}\|e| - K_{Z_3}|\tilde{T}| \|\tilde{Z}\|_F^2$$

where $C_0 = \varepsilon_N + 2Z_M\sqrt{L}$, $C_1 = \sqrt{L}(c_0 + \Theta_d)$.

Taking $K_{Z_1} > \sqrt{L}$, $K_{Z_2} > 1$, and $K_{Z_3} > c_1\sqrt{L}$ yields

$$\dot{L} \leq -(K_f + B)|e|^2 + kZ_M|\tilde{T}| \|\tilde{Z}\|_F - k|\tilde{T}| \|\tilde{Z}\|_F^2 - K_b|\tilde{T}|^2$$
$$+ C_0|\tilde{T}| + C_1|\tilde{T}| \|\tilde{Z}\|_F \quad (44)$$

$$\dot{L} \leq -(K_f + B)|e|^2 - |\tilde{T}|[K_b|\tilde{T}| + k\|\tilde{Z}\|_F^2 - (kZ_M + C_1)\|\tilde{Z}\|_F - C_0]. \quad (45)$$

Completing the square yields

$$\dot{L} \leq -(K_f + B)|e|^2 - |\tilde{T}|[K_b|\tilde{T}| + k\{\|\tilde{Z}\|_F - (\frac{Z_Mk + C_1}{2k})\}^2 - k(\frac{Z_Mk + C_1}{2k})^2 \quad (46)$$
$$- C_0]$$

Thus, the $\dot{L}$ is negative as long as

$$|\tilde{T}| > \frac{k(\frac{Z_Mk + C_1}{2k})^2 + C_0}{K_b}, \quad (47)$$

or

$$\|\tilde{Z}\|_F > \frac{Z_Mk + C_1}{2k} + \sqrt{(\frac{Z_Mk + C_1}{2k})^2 + \frac{C_0}{k}}. \quad (48)$$

According to standard Layapnov theorem, If the error $\tilde{T}$ is greater than the right side of (47), the error is reduced. Equation (49) gives a practical bound on the error

$$|\tilde{T}| \leq \frac{k(\frac{Z_Mk + C_1}{2k})^2 + C_0}{K_b}. \quad (49)$$

Similarly, Eq. (48) gives

$$\|\tilde{Z}\|_F \leq \frac{Z_Mk + C_1}{2k} + \sqrt{(\frac{Z_Mk + C_1}{2k})^2 + \frac{C_0}{k}}. \quad (50)$$

By increasing the gain $K_b$, the stability radius can be reduced to some extent. It is noted that PI controller without NN compensation requires much higher gain to achieve the same performance. In addition, it is difficult to ensure the stability of such highly nonlinear system using only PI controllers. With the NN compensation, the stability of the system is proven, and the tracking error can be arbitrarily kept small by increasing the gain $K_b$. The NN weight errors are basically limited in terms of $V_M$, $W_M$.

It is simple to initialize the NN weights because of the form of a feedforward compensator with a unity feedforward path and an NN parallel path. The initial weights $V$ are selected randomly, while the initial weights $W$ are to set zero. The PI loop with an unity gain feedforward path then maintains the system stable until the NN starts learning.

## 5. Simulation Results

In this section, the author explains the effectiveness of the NN compensator through computer simulations. One consider the system [22] and piecewise linear nonlinearities:

$$M = 0.015, \quad B = 0.951, \tag{51}$$

and a piecewise linear nonlinearity characteristic $P(m_{r1}, m_{r2}, m_{l1}, m_{l2}, u_r, u_l; \cdot)$ where $m_{r1} = 1.0$, $m_{r2} = 2.0$, $m_{l1} = 0.7$, $m_{l2} = 0.5$, $u_r = 0.7$, $u_l = -0.6$. The NN weight tuning parameters are chosen as $S = 8I_9$, $Q = 9I_5$, $k = 0.002$, when $I_N$ is $N \times N$ identity matrix. The robustifying signal gains are $K_{Z_1} = 5.2$, $K_{Z_2} = 3.5$, $K_{Z_3} = 5.8$. The controller parameter $K_p = 0.3$, $K_I = 1.1$, $K_b = 0.4$. The NN has $L = 8$ hidden layer nodes. The inputs to hidden layer weights $V$ are initialized randomly. It is evenly and randomly distributed between 1 and -1. The hidden to output layer weights $W$ are initialized at zero. This weight initialization does not affect system stability because the weight $W$ is initialized at 0, so there is no initial input to the system except for the PI loop. Filter that generates the signal $\dot{T}_{des}$ is implemented as $\dfrac{s}{s+100}$. In. Fig. 6, the tracking performance of the closed-loop system with or without the piecewise linear nonlinearities compensation is shown. The piecewise linear nonlinearities degrades the system performance, since it causes the lost of information about signal $u(t)$, whenever the signal $u(t)$ changes its operation region. Applying the NN compensator greatly reduces the tracking error in Fig. 7. It should be noted that the NN compensator does not handle steady-state errors remaining in the system after the PI controller is operated. It is up to the feedback controller(e.q. PI controller) to take care of it and the NN compensator will only compensate for piecewise linear nonlinearities. The control input with NN compensation $T(t)$ in Fig. 8, except oscillation, converge to the control input without piecewise linear nonlinearity, which means the desired function of the piecewise linear nonlinearities compensation. The proposed NN compensation through simulation is an efficient method of compensating for piecewise linear nonlinearity.

## 6. Conclusions

A new technique for the NN compensation has been proposed for systems. The compensator scheme has a dynamic inversion structure with NN in the feedforward path approximating the piecewise linear nonlinearities inversion error and filter dynamics required for backstepping design. Using nonlinear stability techniques, the bound on tracking errors are derived from tracking error dynamics. According to the simulation results, system performance may be greatly improved by the NN compensation system.

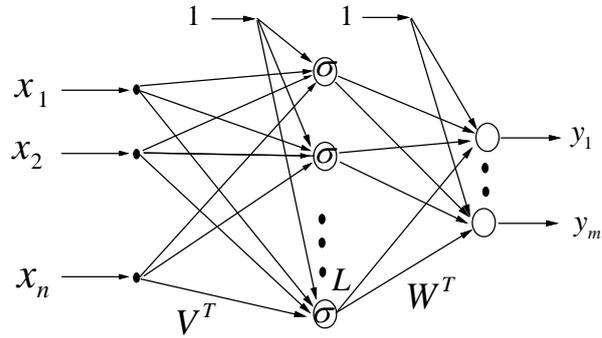

Fig. 1 Three layer Neural Networks.

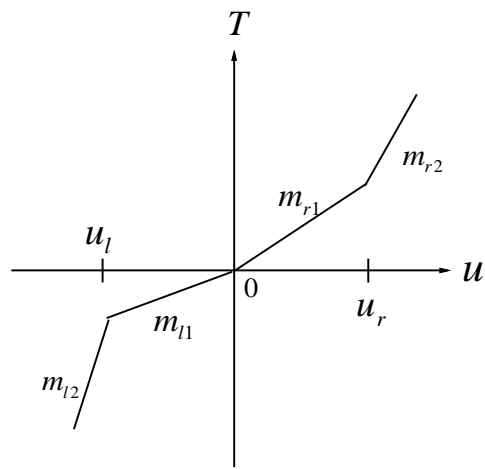

Fig. 2. Piecewise linear nonlinearity.

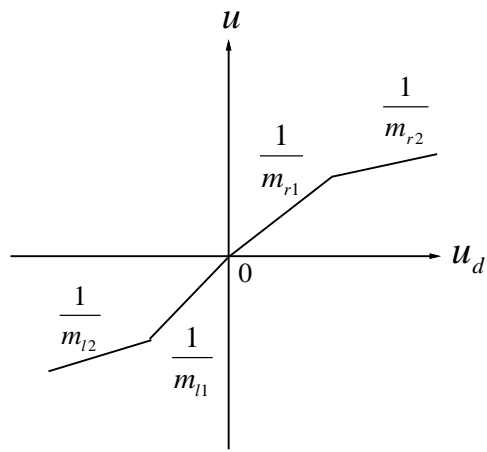

Fig. 3. Inverse of piecewise linear nonlinearity

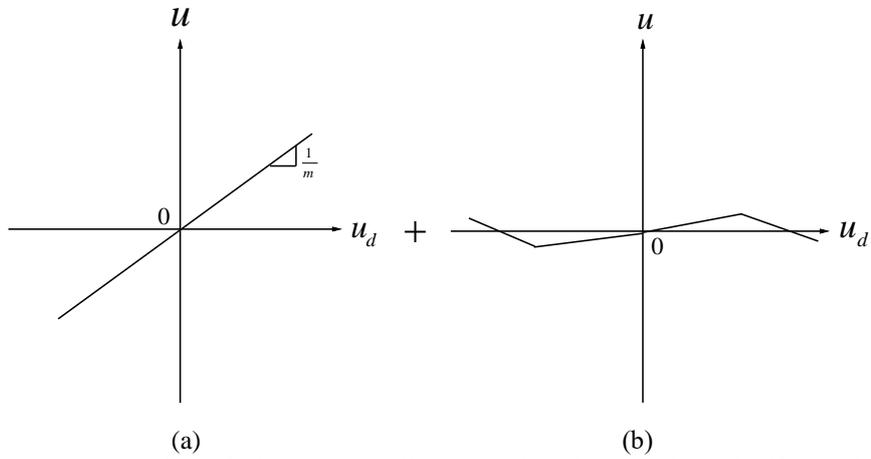

(a) (b)

Fig. 4. Inverse decomposition of piecewise nonlinear nonlinearity (a) direct feedforward term (b) modified hysteresis inverse term.

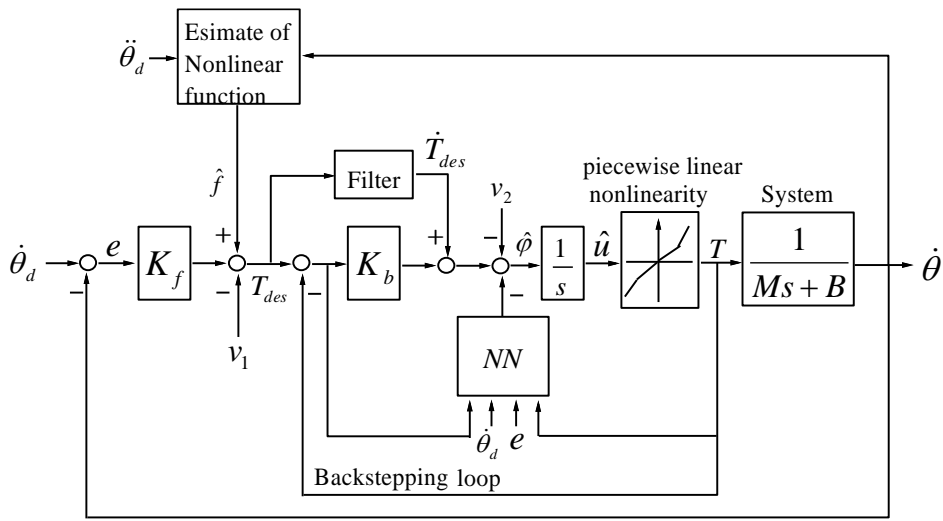

Fig.5 NN piecewise linear nonlinearity compensation of systems

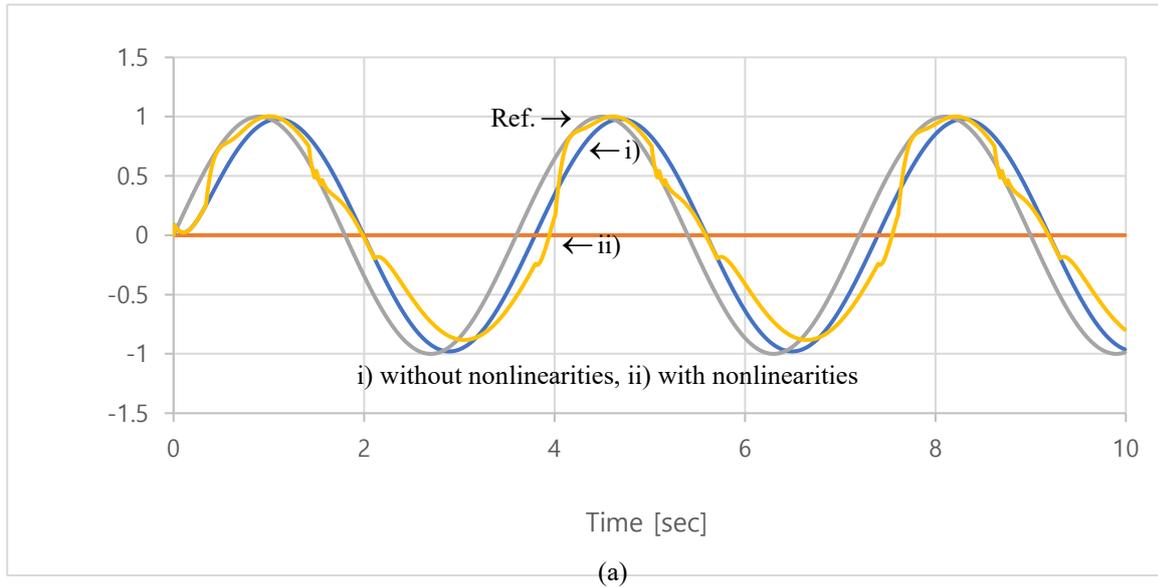

(a)

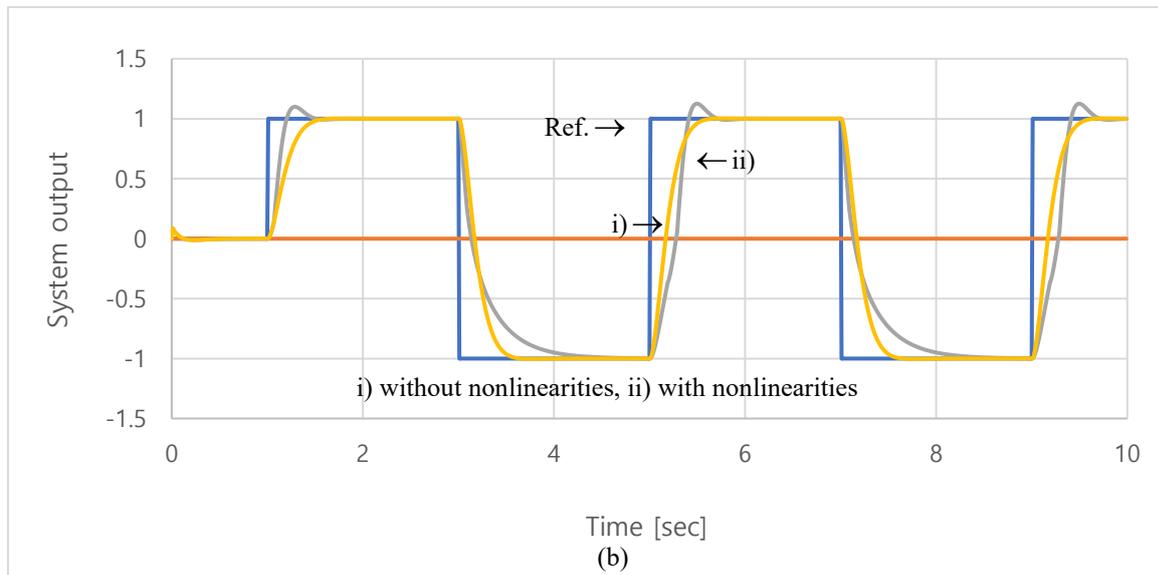

(b)

Fig. 6. System output with/without piecewise linear nonlinearities (a) sinusoidal reference signal (b) rectangular reference signal.

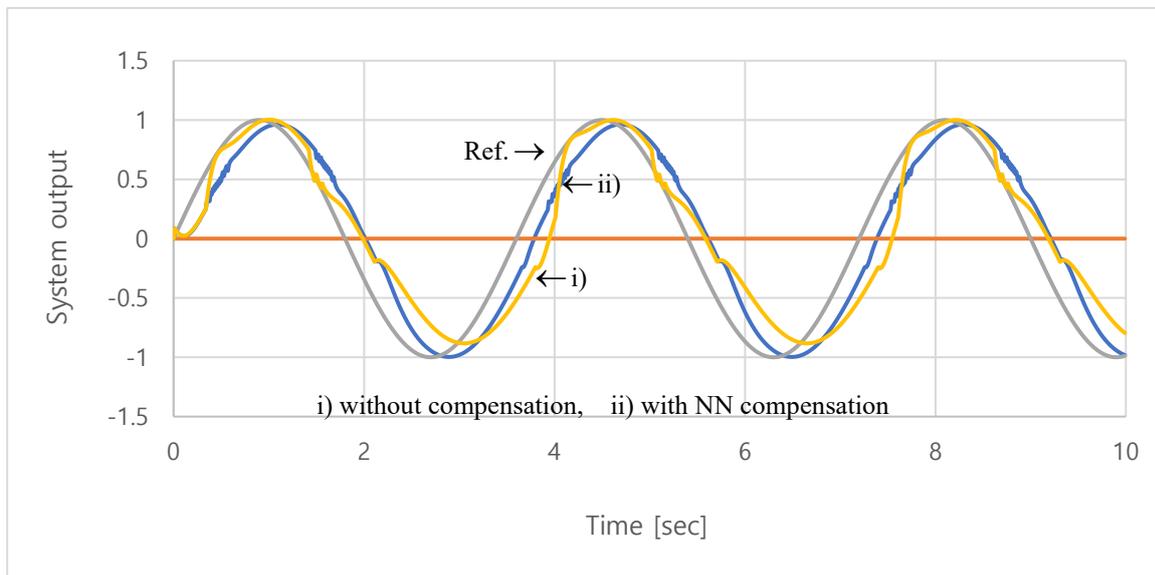

(a)

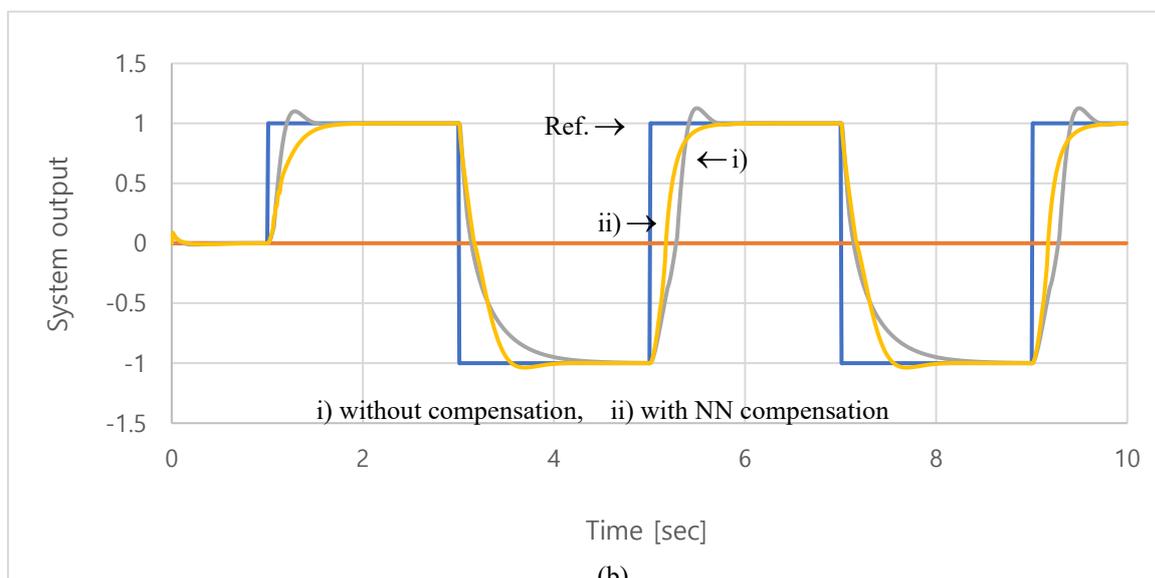

(b)

Fig. 7. System output with/without NN compensation (a) sinusoidal reference signal (b) rectangular reference signal.

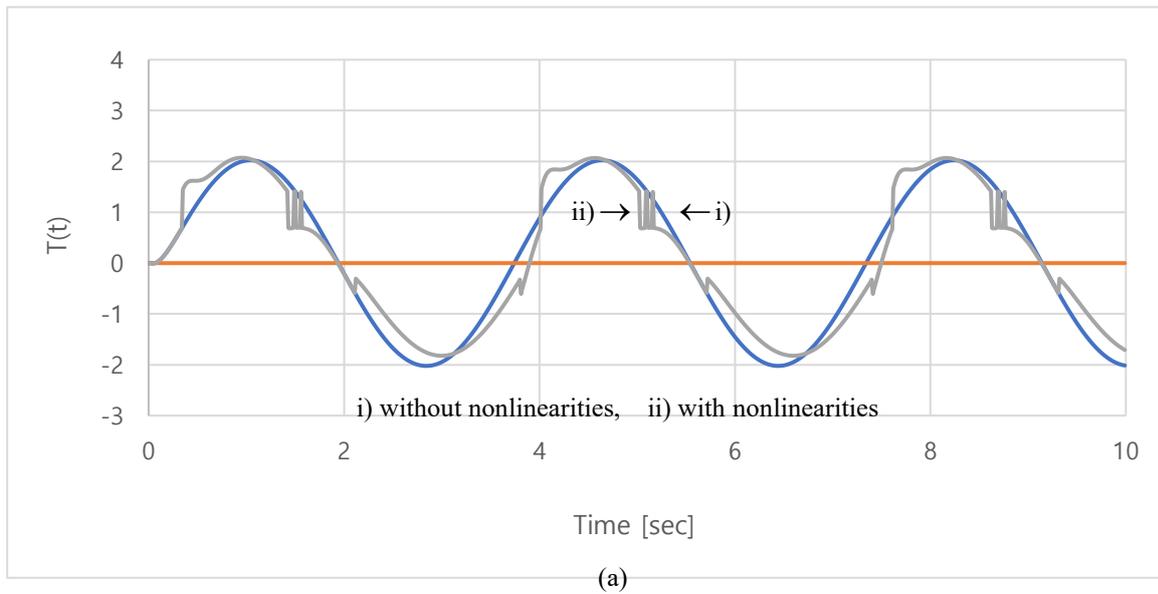

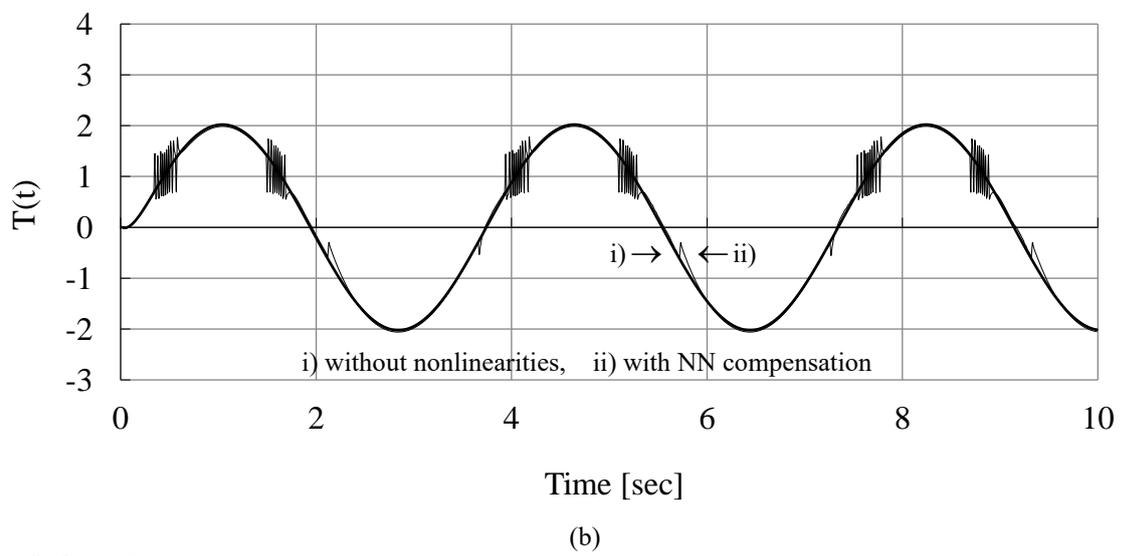

Fig. 8. Control inputs.